\pdfoutput=1

\documentclass[journal]{IEEEtran}

\usepackage[pdftex]{graphicx}
\usepackage{amsmath}
\usepackage{algorithmic}
\usepackage{siunitx}
\usepackage{authblk}
\usepackage[bookmarks=true,breaklinks=true,letterpaper=true,colorlinks,citecolor=blue,linkcolor=blue,urlcolor=blue]{hyperref}
\usepackage{booktabs}
\usepackage{multirow}
\usepackage{fancyhdr}
\usepackage{cite}
\usepackage[many]{tcolorbox}

\DeclareSIUnit\sample{Sample}

\ifCLASSOPTIONcompsoc
  \usepackage[caption=false,font=normalsize,labelfont=sf,textfont=sf]{subfig}
\else
  \usepackage[caption=false,font=footnotesize]{subfig}
\fi

\usepackage{ifthen}
\usepackage{amssymb}
\usepackage{xcolor}

\newboolean{showcomments}
\setboolean{showcomments}{true}

\makeatletter
\newcommand{\mynote}[3]{%
  \ifthenelse{\boolean{showcomments}}{%
   \fbox{\bfseries\sffamily\scriptsize#1}%
   {\small$\blacktriangleright$\textsf{\emph{\color{#3}{#2}}}$\blacktriangleleft$}}%
  {%
   \@bsphack
   \@esphack
  }%
}
\makeatother

\setboolean{showcomments}{true}

\begin{document}
\renewcommand{\arraystretch}{1.1}

\title{A 65\,nm Bayesian Neural Network Accelerator with 360\,fJ/Sample In-Word
GRNG for AI Uncertainty Estimation}

\author[1]{Zephan M. Enciso}
\author[2]{Boyang Cheng}
\author[2]{Likai Pei}
\author[2]{Jianbo Liu}
\author[2]{Steven Davis}
\author[1]{Michael Niemier}
\author[2]{Ningyuan Cao}

\affil[1]{Department of Computer Science and Engineering\\University of Notre Dame}
\affil[2]{Department of Electrical Engineering\\University of Notre Dame}

\maketitle

\begin{abstract}

	Uncertainty estimation is an indispensable capability for AI-enabled,
	safety-critical applications, e.g. autonomous vehicles or medical diagnosis.
	Bayesian neural networks (BNNs) use Bayesian statistics to provide both
	classification predictions and uncertainty estimation, but they suffer from
	high computational overhead associated with random number generation and
	repeated sample iterations.  Furthermore, BNNs are not immediately amenable
	to acceleration through compute-in-memory architectures due to the frequent
	memory writes necessary after each RNG operation. To address these
	challenges, we present an ASIC that integrates 360\,fJ/Sample Gaussian RNG
	directly into the SRAM memory words.  This integration reduces RNG overhead
	and enables fully-parallel compute-in-memory operations for BNNs.  The
	prototype chip achieves 5.12\,GSa/s RNG throughput and 102\,GOp/s neural
	network throughput while occupying 0.45\,mm$\mathbf{^2}$, bringing AI
	uncertainty estimation to edge computation.

\end{abstract}

\definecolor{sigma}{HTML}{c01c28}
\definecolor{mu}{HTML}{613583}

\section{Introduction}
\label{sec:intro}

\begin{figure}[t]
	\centering
	\includegraphics[width=\columnwidth]{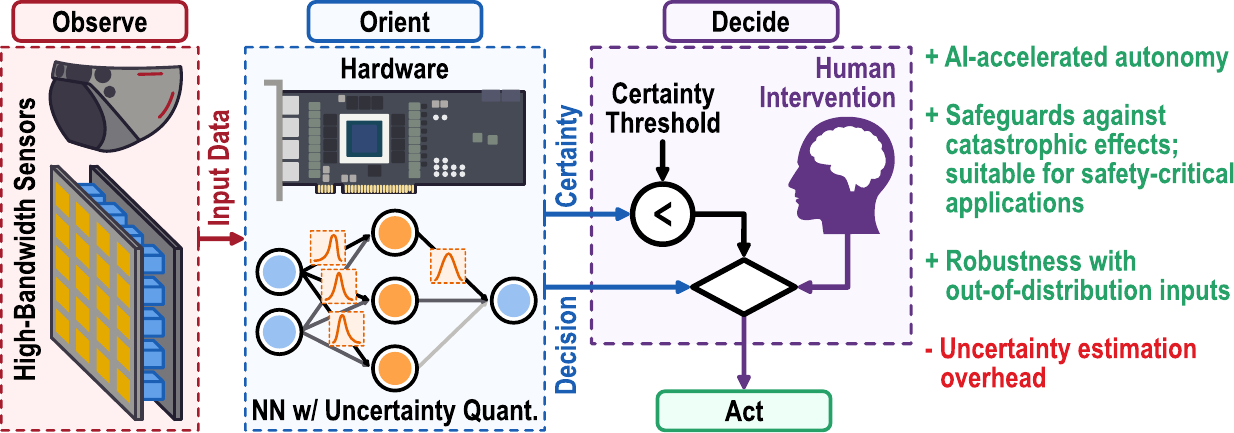}

	\caption{The role of uncertainty estimation in safety-critical
	applications. During typical operation, the model rapidly performs
	autonomous actions. However, if the model's certainty falls below a
	predetermined threshold, it would request human intervention to avoid
	catastrophic effects.}
	\label{fig:motivation}

\end{figure}

Uncertainty estimation is crucial for robust decision-making in data-driven deep
learning (DL) systems, particularly when these systems interact with the
physical environment. In safety-critical applications, such as autonomous
vehicle navigation and obstacle avoidance, medical diagnosis, aerospace control
systems, and industrial automation, models equipped with uncertainty estimation
could trigger human intervention or engage alternative sensors and models when
prediction confidence drops below a set threshold.  This process significantly
mitiages the risk of potential catastrophic outcomes (see
Fig.~\ref{fig:motivation}). 

\begin{figure}[b]
	\centering
	\includegraphics[width=\columnwidth]{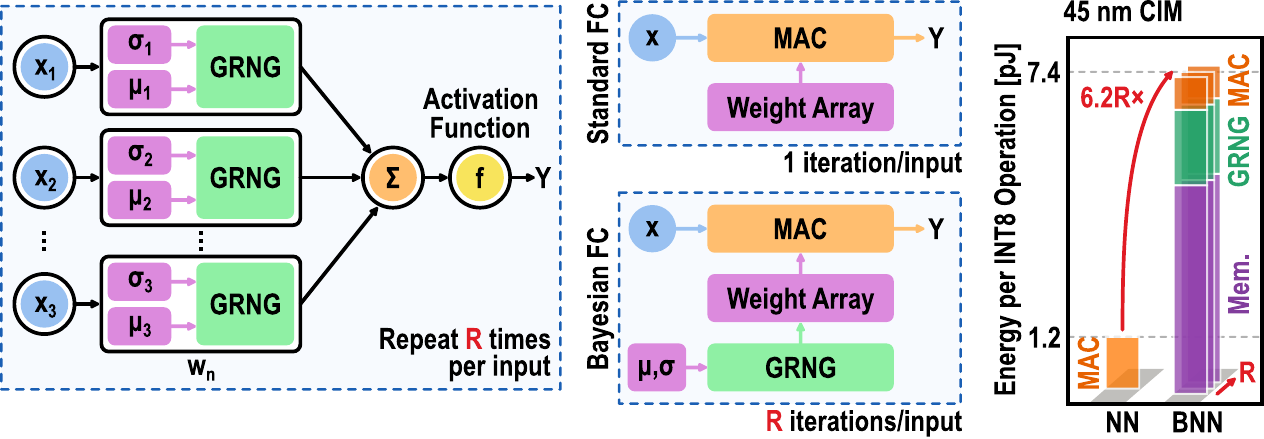}

	\caption{\textbf{Left)} Conventional BNN neuron.  Each weight uses GRNG to
	sample from a Guassian distribution, so the weight must store distribution
	properties $\mu$ and $\sigma$.  \textbf{Right)} BNN fully-connected (FC) layers
	incur significant overhead from multiple memory operations and GRNG
	compared to standard FC layers.}
	\label{fig:overhead}

\end{figure}

Bayesian neural networks (BNNs) provide a DL framework capable of delivering
probabilistic estimates of classification uncertainty by replacing conventional
deterministic weights with a posterior distribution of weights
\cite{shridhar2019bayesian, gawlikowski2023survey}. Deployed models typically
approximate the posterior with Gaussian distributions \cite{blei2017variational,
hoffman2013stochastic}; even so, BNNs incur significant overheads from Gaussian
random number generation (GRNG), the associated memory accesses, and repeated
inferences, as shown in Fig.~\ref{fig:overhead}.

Moreover, BNNs derive less benefit from non-Von Neuman architectures, such as
compute-in-memory (CIM) \cite{khaddam2022hermes, wu202228nm}. When BNNs are
deployed on these architectures, the GRNG must retrieve distribution parameters
from memory, generate a weight sample, and subsequently write the sample back to
the CIM array. Simulations indicate that even CIM-accelerated BNNs consume more
than six times the energy per \texttt{INT8} operation in each sampling iteration
compared to traditional neural networks \cite{verma2019memory, horowitz20141}
(see Fig.~\ref{fig:overhead}), increasing the cost of deploying them on edge
inference engines.

Current BNN accelerators focus primarily on either enhancing the efficiency of
GRNG hardware \cite{dorrance2023energy, shukla2021ultralow, cai2018vibnn,
xu2021bayesian, fan2022fpga, lee2005hardware, thomas2013multiplierless} or
maximizing data reuse \cite{fan2022accelerating}. By contrast, this chip
performs fully parallel, in-memory matrix-vector multiplication with arbitrary
Gaussian weight distributions.  Crucially, this is achieved without requiring
extra memory accesses for the GRNG, as the GRNG is integrated within the memory
words. This stochastic, mixed-signal CIM architecture, combined with
state-of-the-art (SOTA) GRNG energy and area efficiency, enables energy-- and
area--efficient BNN acceleration.

\section{Background}
\label{sec:background}

\subsection{Bayesian Neural Networks}
\label{sec:bnn}

Bayesian neural networks (BNNs) augment traditional neural networks by encoding
weights and biases as posterior distributions, which facilitates a probabilistic
interpretation of model predictions~\cite{goan2020bayesian}. This probabilistic
approach enables BNNs to provide both classification predictions and uncertainty
estimations. 

Formally, the posterior distribution of weights follows Bayes' Theorem:

\begin{equation}
	P\left(\mathbf{W}\, |\, \mathbf{X}, \mathbf{Y}\right) = \frac{P\left(\mathbf{Y}\, |\, \mathbf{W}, \mathbf{X}\right) P(\mathbf{W})}{P(\mathbf{X})}
\end{equation}

\noindent where:

\begin{itemize}

\item $P\left(\mathbf{W}\, |\, \mathbf{X}, \mathbf{Y}\right)$ denotes the
	posterior probability of weights $\mathbf{W}$ given input $\mathbf{X}$ and
	output $\mathbf{Y}$.

\item $P\left(\mathbf{Y}\, |\, \mathbf{W}, \mathbf{X}\right)$ is the conditional
	probability or likelihood of observing outputs $\mathbf{Y}$ for weights
	$\mathbf{W}$ and input $\mathbf{X}$.

\item $P(\mathbf{W})$ and $P(\mathbf{X})$ are the prior probabilities of the
	weights and input, respectively.

\end{itemize}

Directly approximating the posterior distribution $P\left(\mathbf{W}\, |\,
\mathbf{X}, \mathbf{Y}\right)$ is computationally intractable for edge devices.
Consequently, deployed models typically approximate the posterior with a
Guassian distribution through a method known as variational inference (VI)
\cite{hoffman2013stochastic}.  This approximation is expressed as:

\begin{equation}
	P\left(\mathbf{W}\, |\, \mathbf{X}, \mathbf{Y}\right) \approx \mathcal{N} \left( \mathbf{W}\, |\, \mathbf{\mu}, \mathbf{\sigma} \right)
\end{equation}

\noindent where:

\begin{itemize}

\item $\mathbf{\mu}$ is the mean of the Gaussian distribution.

\item $\mathbf{\sigma}$ is the covariance matrix of the Gaussian distribution.

\end{itemize}

The process of approximating the posterior distribution involves minimizing the
divergence between the true posterior and the approximated Gaussian
distribution.  This minimization is commonly achieved by maximizing the evidence
lower bound (ELBO), a technique that balances the fit of the model to the data
with the complexity of the model \cite{blei2017variational} .  

\subsection{Compute-in-Memory Accelerators}
\label{sec:cim}

Compute-in-memory (CIM) is an emerging accelerator architecture designed to
address the von Neumman bottleneck, which refers to the significant latency and
power consumption resulting from the separation of memory and processing units
in traditional computing architectures.  By integrating memory and computation,
CIM accelerators aim to reduce data movement, thereby improving both energy
efficiency and performance~\cite{verma2019memory, yu2021compute}.  This
integration is particularly beneficial for DL applications, where the energy
cost and latency of frequent memory access can be substantial and memory access
patterns are relatively simple.

One of the key principles behind CIM is the use of crossbar arrays for
performing matrix-vector multiplications (MVMs) directly within the memory
\cite{zhang2018neuromorphic}. In a crossbar array, the matrix weights are
encoded as the conductance $\mathbf{G}_{(i,j)}$ of each memory cell.  When a
voltage $\mathbf{X}_i$ is applied to each row, it induces a current
$\mathbf{Y}_j$ to flow down each column according to Kirchhoff's Law.
Mathematically, this can be represented as:

\begin{equation}
	\mathbf{Y}_j = \sum_i^N \mathbf{X}_i \mathbf{G}_{(i,j)}
\end{equation}

Thus, $\mathbf{Y} = \mathbf{G} \cdot \mathbf{X}$, and this direct in-memory
computation reduces the need for data transfer between memory and computation
units.

Crossbars also support parallel in-memory operations. Depending on the precision
and number of downstream analog-to-digital converters (ADCs), multiple rows and
columns may be activated concurrently. This parallelism enables CIM accelerators
to perform high-throughput computation, which is essential for DL applications
that require extensive MVMs.  CIM designs have been realized with both static
random-access memory (SRAM) and emerging memory technologies (EMTs), such as
resistive random-access memory (ReRAM), phase-change memory (PCM), and
spin-transfer torque magnetic random-access memory (STT-MRAM)
~\cite{jhang2021challenges, li2019long, jung2022crossbar}.  These technologies
may offer advantages in terms of storage density and non-volatility, making them
suitable for different application requirements.

\subsection{BNN Hardware Acceleration}

BNNs are distinguished from conventional neural networks by their reliance on
stochastic sampling, which significantly amplifies hardware resource demands.
This stochastic process necessitates extensive inference runs to accurately
determine the mean and variance of inference scores, thereby assessing model
uncertainty. Digital BNN accelerators focus on optimizing
GRNG~\cite{thomas2013multiplierless, lee2005hardware} or improving the hardware
architecture pipeline through data reuse strategies that minimize unnecessary
data transactions~\cite{fan2022accelerating, xu2021bayesian}.  Despite these
efforts, an efficiency gap remains due to the limitations of digital GRNG and
the frequency of memory operations required for BNN inference.

Each inference iteration in a BNN involves reading distribution parameters,
generating a Gaussian sample, and subsequently updating the weight array.  This
iterative process makes it challenging to apply CIM architectures for BNN
hardware acceleration. Leveraging the stochastic properties of EMTs, can enable
efficient GRNG through device variation ~\cite{bonnet2023bringing}.  These
emerging memory technologies exploit the inherent stochasticity in their
physical properties to generate noise with the stored data, thereby integrating
GRNG with memory.

However, the dependency on specific device characteristics for generating
stochasticity, coupled with the high power consumption of memory writes for
storing GRNG results, introduces programming complexities and scalability
issues.  Additionally, the endurance and durability concerns of these
technologies further complicate their long-term viability for BNN acceleration.

\section{Chip Architecture and Circuit Design}

\subsection{Hardware-Software Co-Design}
\label{sec:arch}

This chip employs several algorithmic optimizations to reduce BNN overhead and
simplify GRNG. First, SOTA Bayesian accelerators typically employ Bayesian
weights only to the last fully-connected (FC) layers (``partial BNN''). These
layers are critical for generating a classification from previously-extracted
features and thus have the largest impact on uncertainty. By doing so, partial
BNNs significantly reduce the number of repeated operations and the requisite
number of RNG samples.  Meanwhile, the computationally-expensive convolutional
and/or recurrent layers are processed as standard, non-Bayesian layers.  This
strategy maintains the model's ability to quantify uncertainty without incurring
excessive computational costs \cite{fan2022fpga}.

A key optimization in this architecture is weight decomposition, which
separates each weight into a sum of the mean $\mu$ and the product of the its
standard deviation $\sigma$ and a distribution sample $\epsilon$:

\begin{equation}
	w_{\left(i,j\right)} = \mu_{\left(i,j\right)} + \sigma_{\left(i,j\right)}\epsilon,\ \epsilon \sim \mathcal{N}\left(0, 1\right)
\end{equation}

Thus, the $j^\text{th}$ output $\mathbf{Y}_j$ with $N$ inputs
$\mathbf{X}_{\left(i,j\right)}$ is given by:

\begin{equation}
	\mathbf{Y}_j = f \left( \sum_{i=1}^N \mathbf{X}_{\left(i,j\right)}\mu_{\left(i,j\right)} + \sum_{i=1}^N \mathbf{X}_{\left(i,j\right)}\sigma_{\left(i,j\right)}\epsilon \right)
\end{equation}

Since the mean $\mu$ is static, it only needs to be processed once. Furthermore,
$\epsilon$ is sampled from the same standard normal distribution instead of a
parameterized normal distribution, which significantly simplifies the GRNG
design.  This simplification is crucial because it reduces the computational
burden associated with generating and storing multiple unique random samples for
each weight update.

\subsection{CIM Tile Architecture}
\label{sec:tile}

\begin{figure}[t]
	\centering
	\includegraphics[width=\columnwidth]{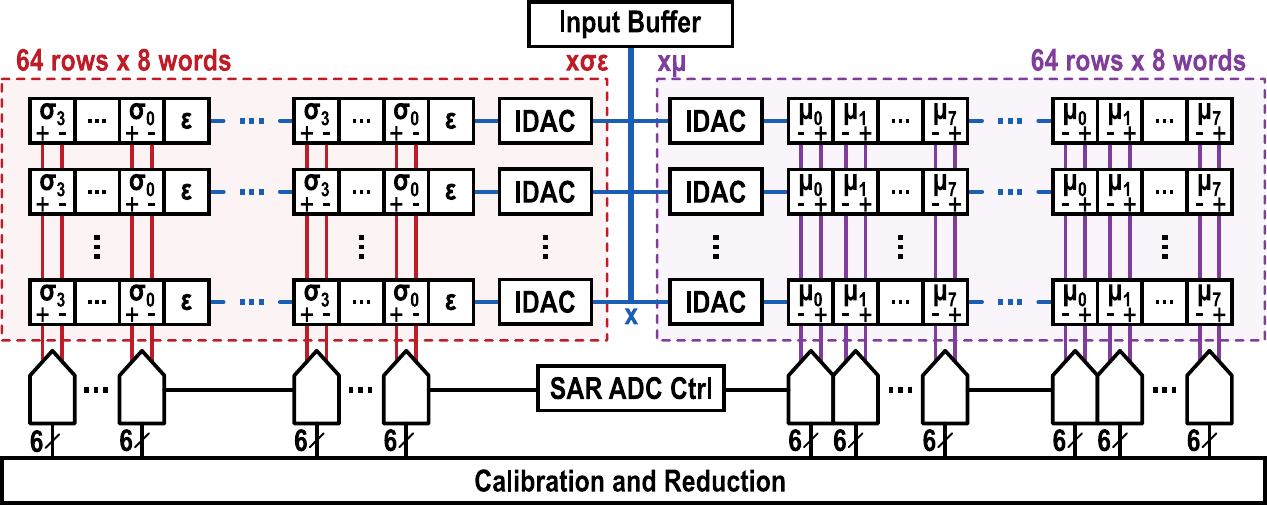}

	\caption{CIM tile architecture, featuring two subarrays for separately
	computing \textcolor{sigma}{$\mathbf{X}\sigma\epsilon$} and
	\textcolor{mu}{$\mathbf{X}\mu$}. Both subarrays receive the same input
	$\mathbf{X}$, and downstream reduction logic recombines the results.}

    \label{fig:arch}

\end{figure}

The CIM tile architecture is designed to reflect the weight decomposition
methodology, as shown in 
Fig.~\ref{fig:arch}, 
by using two crossbar subarrays for separately computing
$\mathbf{X}\mu$ and $\mathbf{X}\sigma\epsilon$. In the fabricated prototype,
each CIM tile comprises 64 rows of 8 words.  Each word consists of a 8-bit $\mu$ and a
4-bit $\sigma$. The 4-bit digital input vector $\mathbf{X}$ is fed to current
digital-to-analog converters (IDACs).
Each word bit is associated with a dedicated 6-bit successive-approximation
register (SAR) ADC. 

To enhance efficiency, the SAR ADCs share a common synchronous controller. This
shared control reduces the overall area requirement for each ADC and enables
pitch-matching with the SRAM arrays.  Pitch-matching is critical as it
eliminates the need for column multiplexing, thus enabling single-cycle MVM.

The CIM tile also includes digital reduction logic, which also corrects for
individual ADC offset. This logic shifts and adds the outputs from the SAR ADCs
to reconstruct each word.  Once the individual components of $\mathbf{X}\mu$ and
$\mathbf{X}\sigma\epsilon$ have been computed, they are combined to generate a
single output vector.

\subsection{In-Word GRNG Circuit}
\label{sec:grng}

\subsubsection{Capacitor Thermal Noise}

Unlike pseudorandom RNGs, true RNGs (TRNG) require a physical process capable of
producing entropy~\cite{turan2018recommendation}.  Common entropy sources for
integrated circuits include thermal noise, jitter, and electrical
metastability~\cite{sunar2009true}. \cite{taneja2021memory} identified that
thermal noise affects the discharge period of a capacitor with constant leakage
current, and the discharge follows a Gaussian distribution whose properties are
given by:

\begin{align}
	\mu_T &= \frac{C V_{DD}}{2 I_L} \\
	\sigma^2_T &= \frac{\mu_T q}{2 I_L}
\end{align}

\noindent where:

\begin{itemize}

\item $\mu_T$ is the mean or expectation of the distribution.
	
\item $\sigma^2$ is the variance of the distribution.

\item $C$ is the capacitance of the capacitor being discharged.

\item $I_L$ is the total leakage current.

\item $q$ is the initial charge on the capacitor.

\item $T$ represents the time after the capacitor begins leaking that the
	voltage crosses threshold $V_{Thr}$.

\end{itemize}

\subsubsection{GRNG Operation}

\begin{figure}
	\centering
	\includegraphics[width=\columnwidth]{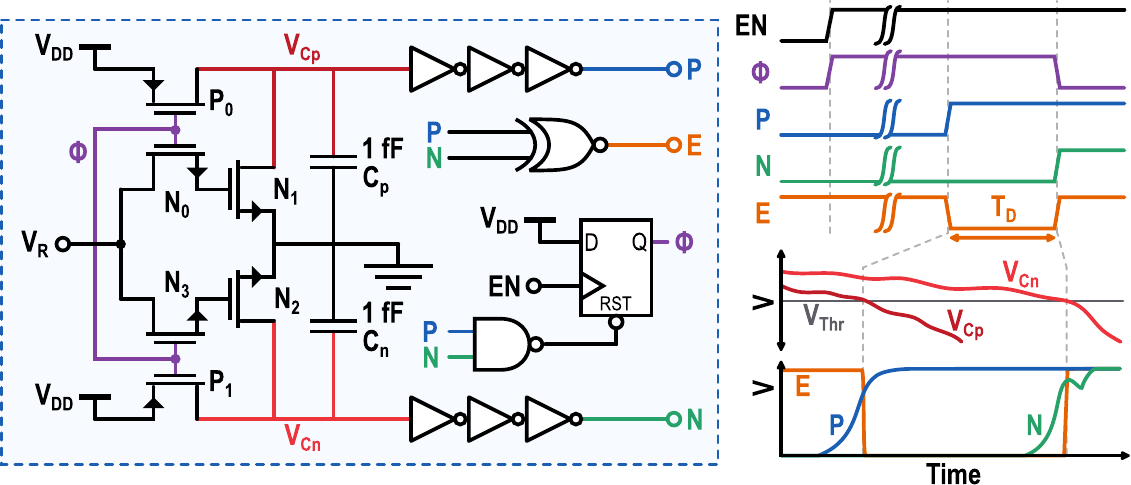}

	\caption{GRNG circuit and timing diagram.  Thermal noise causes $C_n$ to
	discharge at a different rate than $C_p$, producing an ouput pulse
	\texttt{E} whose duration follows a 0--mean Gaussian distribution.}
	\label{fig:grng}

\end{figure}

The in-word GRNG circuit (see Fig.~\ref{fig:grng}) that enables CIM-accelerated
BNNs compares the discharge time of two capacitors $C_p$ and $C_n$ to yield a
normal distribution centered on zero.  The process involves the following steps:

\begin{enumerate}

\item Initially $P_0$ and $P_1$ charge $C_p$ and $C_n$ to $V_{DD}$ while $\Phi$
	is low.  

\item Pulling \texttt{EN} high latches $\Phi$ and causes $N_1$ and
	$N_2$---biased to $V_R$ through $N_0$ and $N_3$, respectively---to slowly
	discharge $C_p$ and $C_n$ to ground.  

\item The inverters at nodes $V_{Cp}$ and $V_{Cn}$ sharpen the transition over
	inverter threshold $V_{Thr}$, yielding digital signals \texttt{P} and
	\texttt{N}, which represent a positive and negative value, respectively.  

\item The output \texttt{E} is the logical \texttt{XNOR} of \texttt{P} and
	\texttt{N}, and \texttt{E} represents the sampled random variable encoded in
	the time domain.  $T_D$, the width of output pulse \texttt{E}, follows a
	Gaussian distribution; Sec.~\ref{sec:grngeval} validates the normality
	of the output distribution with measured results.

\end{enumerate}

The capacitors $C_p$ and $C_n$ are designed to be small
($\sim$\qty{1}{\femto\farad}) to minimize discharge energy. They are
physically implemented as metal fringe capacitors directly above the
GRNG circuit for optimal area utilization and mismatch performance
\cite{tripathi2014mismatch}.  The inverters dissipate the majority of
the GRNG power because they create a weak conduction path from $V_{DD}$
to ground as $V_{Cp}$ and $V_{Cn}$ approach $V_{Thr}$. To mitigate
inverter power consumption, the GRNG circuit also includes an
asynchronous reset D flip-flop (DFF), which resets $\Phi$ when
\texttt{E} goes high, charging $V_{Cp}$ and $V_{Cn}$ back to $V_{DD}$
and eliminating the conduction path.

\subsubsection{Calibration for Static Variation}

Transistor mismatch induced during fabrication may cause either $N_1$ or $N_2$
to conduct current faster than the other for the same applied gate voltage.
This static variation manifests as a non-zero mean $\epsilon_0$ in the output
distribution:

\begin{equation}
	\epsilon_0 = V_{DD} \left( \frac{C_p I_{N2} - C_n I_{N1}}{2 I_{N1} I_{N2}} \right) \neq 0
\end{equation}

Subthreshold operation, which is required to produce adequately large standard
deviations, further amplifies the effects of transistor variations
\cite{cheng202465nm}. However, such deviation is static---for a given die, the
same variation will be observed each cycle---and can be systematically corrected
through calibration. First, the chip measures the mean offset $\epsilon_{0,
{\left(i,j\right)}}$ for weight $w_{\left(i,j\right)}$ by writing \texttt{1} to
all $\sigma$ words and multiplying each row by \texttt{1} sequentially.  Then,
the static offset is subtracted from the $\mu$ cell, resulting in calibrated
weight $w'_{\left(i,j\right)}$:

\begin{align}
	w_{\left(i,j\right)} &= \mu_{\left(i,j\right)} + \sigma_{\left(i,j\right)} \left( \epsilon + \epsilon_{0, \left(i,j\right)} \right) \\
	w'_{\left(i,j\right)} &= \mu'_{\left(i,j\right)} + \sigma_{\left(i,j\right)}\epsilon,\ \mu'_{\left(i,j\right)} = \mu_{\left(i,j\right)} - \sigma_{\left(i,j\right)}\epsilon_0
\end{align}

The entire calibration process consumes \qty{3.6}{\nano\joule} and must only
be performed once per chip, though subsequent weight changes must be updated to
include the offset.  

\subsection{CIM Memory Words}
\label{sec:words}

\begin{figure}
	\centering
	\includegraphics[width=\columnwidth]{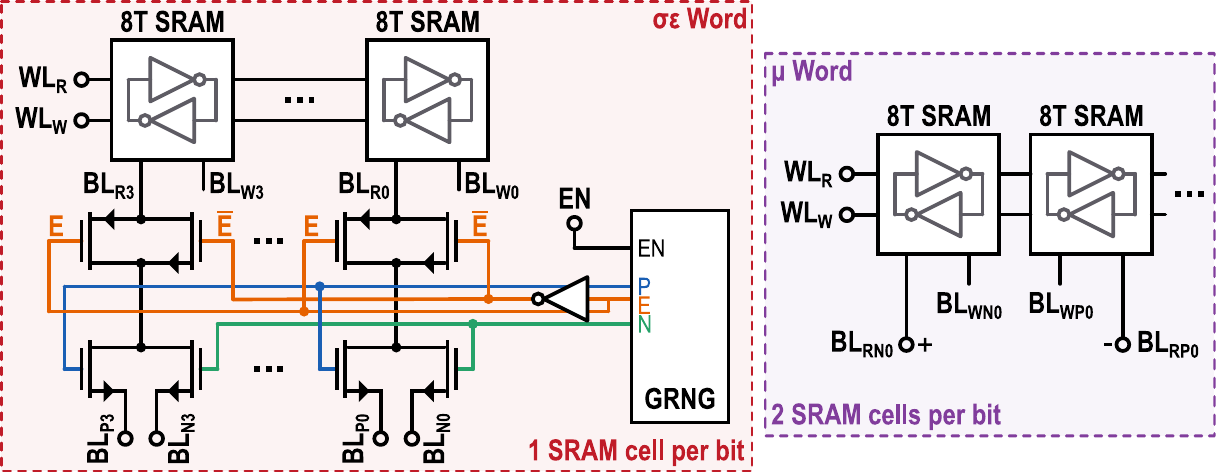}

	\caption{\textcolor{sigma}{$\sigma\epsilon$} and \textcolor{mu}{$\mu$} CIM
	word circuits.  The $\sigma\epsilon$ word contains additional switches to
	interface with the GRNG and produce a differential output.  The $\mu$ word's
	output is differential because the data is stored differentially across 2
	SRAM cells per bit.} 
	\label{fig:sram}

\end{figure}

The CIM memory words, as depicted in Fig.~\ref{fig:sram}, use 8T SRAM cells to
minimize parasitic leakage current.  These cells feature separate wordlines
(WLs) and bitlines (BLs) for reading and writing operations.  The computation
process begins by charging all BLs to $V_{DD}$.  Subsequently, each row's IDAC
converts the 4-bit digital input $\mathbf{X}_i$ into a read WL voltage.  This
voltage ensures that the current conducted by the 8T SRAM cells is linearly
proportional to $\mathbf{X}_i$.  The cells conduct current from the BLs to
ground for a set duration, and the resulting voltage on the $j^\text{th}$ BL
represents the vector dot product of $\mathbf{X}$ by all cells connected to
$BL_j$.

To facilitate robust CIM operation, the downstream SAR ADCs operate
differentially, with charge on $BL_N$ and $BL_P$ representing negative and
positive values, respectively. For differential data encoding in the $\mu$ word,
each bit uses 2 SRAM cells. A positive value is represented by
\texttt{0}\,\texttt{1} and a negative value represented with
\texttt{1}\,\texttt{0}. By contrast, the $\sigma\epsilon$ word only requires one
SRAM cell per bit because the GRNG produces signed outputs.  First, the current
through the SRAM cell is gated by GRNG output pulse \texttt{E} via transmission
gates at the output.  Then, current flows either from $BL_P$ or $BL_N$ depending
on complimentary GRNG signals \texttt{P} and \texttt{N}.

\section{Hardware Evaluation}
\label{sec:eval}

\begin{figure}
	\centering
	\includegraphics[width=\columnwidth]{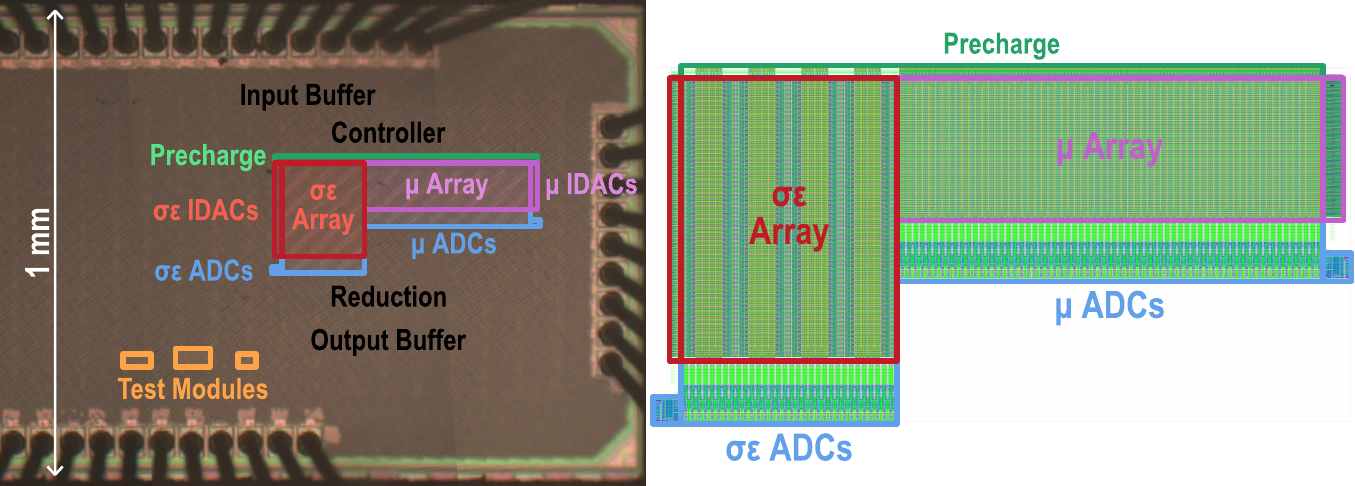}

	\caption{\textbf{Left:} Annotated die shot showing CIM tile (colored
	outlines) and supporting digital logic (black).  The number of IO pads were
	limited to decrease package size and thus decrease bond wire capacitive
	coupling. \textbf{Right:} Detailed view of the CIM tile.}
	\label{fig:dieshot}

\end{figure}

\begin{figure}[h]
	\centering
	\includegraphics[width=\columnwidth]{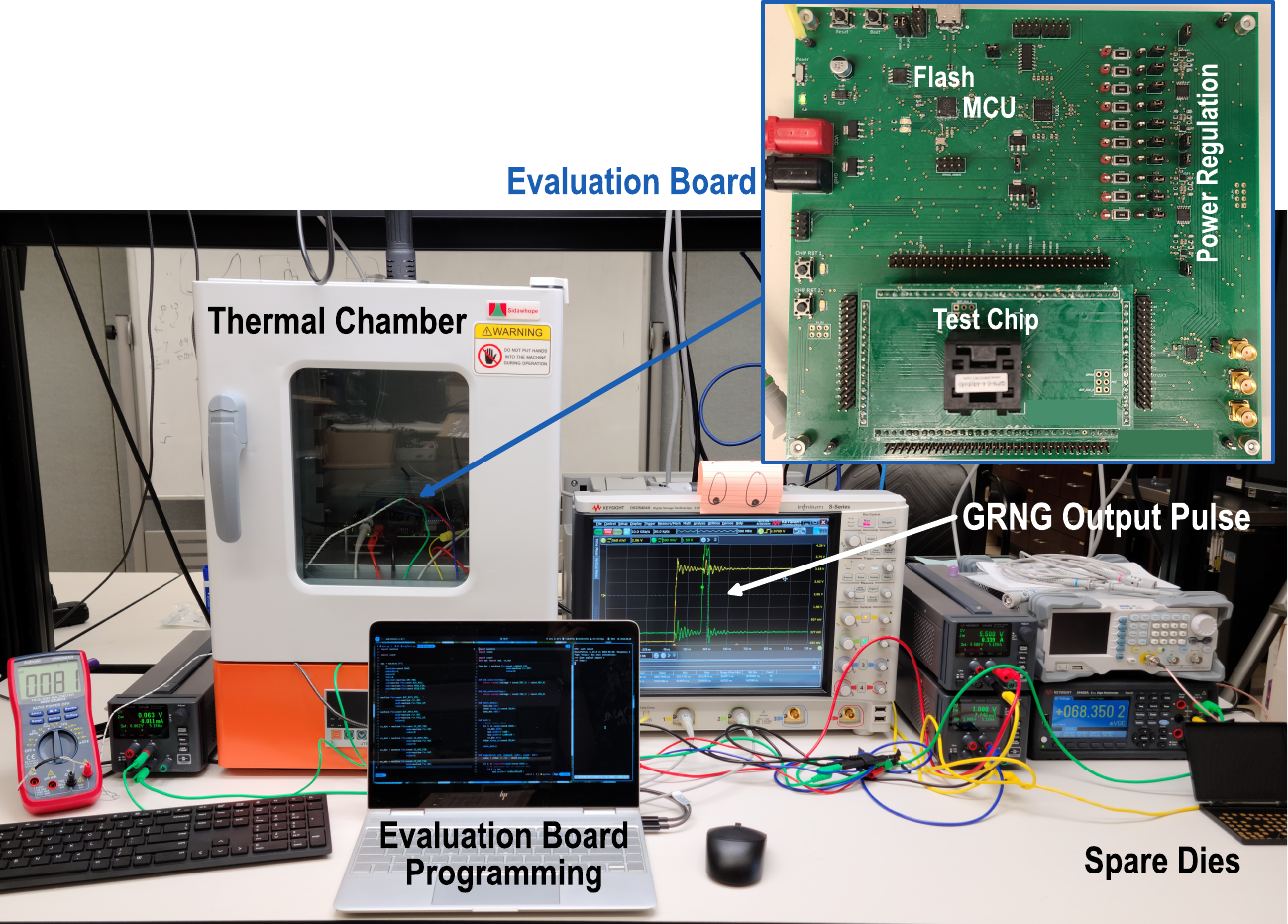}

	\caption{Measurement setup for obtaining GRNG output distributions.  The
	custom evaluation board rests in a temperature-controlled chamber, and a
	DSOS404A oscilloscope records output pulses via a N2795A active probe.}
	\label{fig:measurement}

\end{figure}

A prototype chip fabricated on a commercial 65 nm PDK (Fig.~\ref{fig:dieshot})
provides validation measurements for this design.  GRNG tests were conducted in
a thermal chamber, as shown in Fig.~\ref{fig:measurement} to ensure a stable
operating environment and to measure the temperature stability of the GRNG
circuit.

\subsection{In-Word GRNG}
\label{sec:grngeval}

\begin{figure}
	\centering
	\includegraphics[width=\columnwidth]{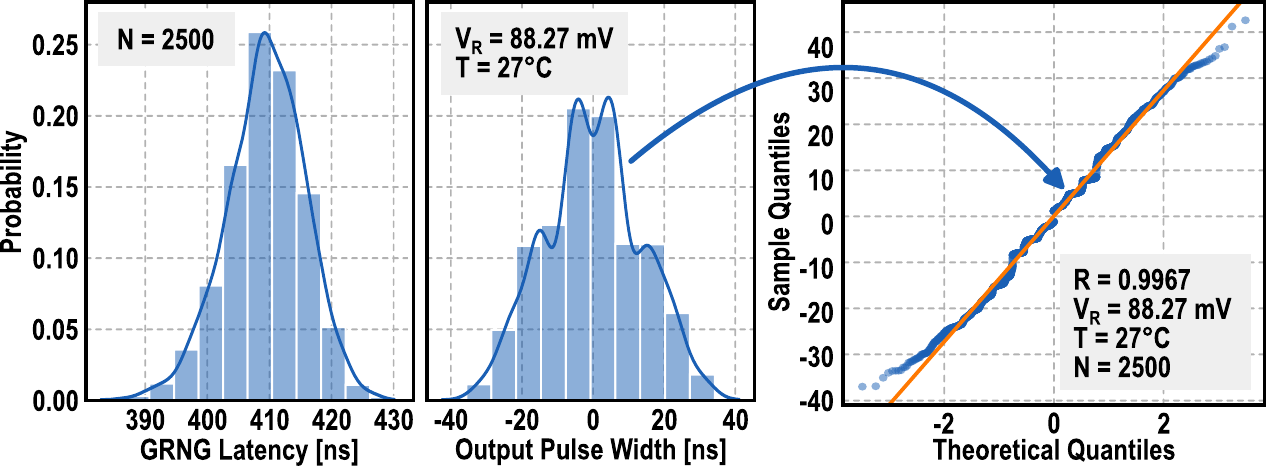}

	\caption{Sample GRNG output pulse width and latency distributions for one
	bias and temperature configuration.  Normal probability plot (Q--Q plot)
	examines normality of pulse width distribution; the orange line represents
	an ideal Gaussian distribution, and the $r$-value reports how well the line
	fits the measured data.  Output pulses less than \qty{1}{\ns} cannot be
	reliably measured due to IO constraints.} 
	\label{fig:distribution}

\end{figure}

The output distribution and associated latency for one bias configuration are
shown in Fig.~\ref{fig:distribution}. An examination of the normal probability
plot---a specific type of Q-Q plot used to assess normal distributions---indicates that
the output distribution is suitably normal for quantized applications with an
$r$-value of 0.9967 for $N = 2500$ samples. 

\begin{figure}
	\centering
	\includegraphics[width=0.75\columnwidth]{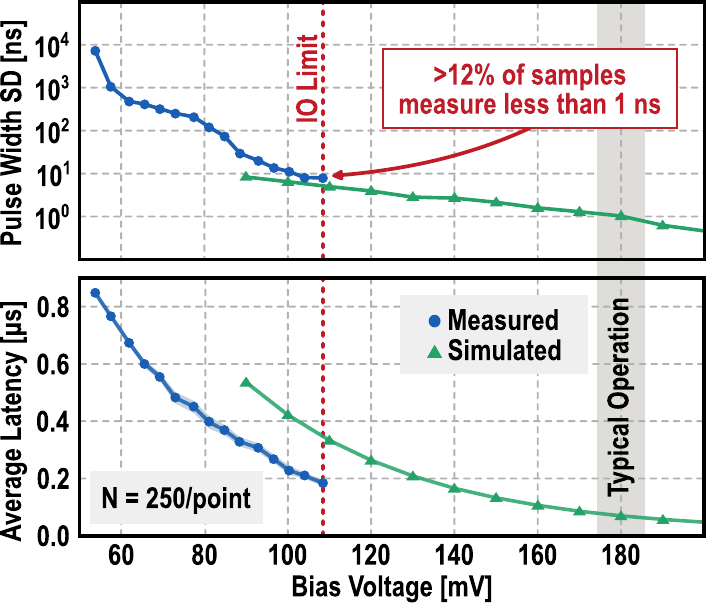}

	\caption{GRNG operation under different bias configurations.  Increasing the
	bias voltage $V_R$ decreases average latency, but it also reduces the output
	standard deviation.  Off-chip measurements of the GRNG output are
	constrained by chip IO, so plots also include simulated datapoints.}
	\label{fig:vr}

\end{figure}

Since the bias voltage $V_R$ controls the rate at which capacitors $N_1$ and
$N_2$ discharge $C_p$ and $C_n$, increasing $V_R$ reduces total energy and GRNG
latency but also decreases the output standard deviation. Fig.~\ref{fig:vr}
illustrates the trade-off between average latency and the standard deviation of
the output pulse width. Typical operation uses a \qty{180}{\milli\volt} bias
voltage to generate an output distribution with a \qty{1.0}{\ns} standard
deviation at an average latency of \qty{69}{\ns}, consuming
\qty{360}{\femto\joule}/Sample.  However, increasing the bias voltage beyond
\qty{110}{\milli\volt} creates an increased proportion of sub-\qty{1}{\ns}
pulses, which compromises measurement accuracy due to the IO limits of the
experimental setup.  Therefore, Fig.~\ref{fig:vr} includes both chip
measurements and parasitic-annotated simulation results to provide a
comprehensive view of GRNG operating points.

Thermal noise is also sensitive to change in environmental temperature; as the
environmental temperature increases from \qty{28}{\celsius} to
\qty{60}{\celsius}, the pulse width standard deviation increases by
$2.62\times$, and the average latency decreases by $2.49\times$ (see
Tab.~\ref{tab:temp}).  Despite these variations, the quality of the output
distribution slightly improves---the normal probability plot $r$-value increases
by $4.78\%$ over the same temperature increase.  Furthermore, changes in
standard deviation can be compensated for by tuning the IDAC bias, which affects
the rate at which the CIM cells conduct current from the bitlines.

\begin{table}[]
\centering
\caption{Measured GRNG temperature stability}
\label{tab:temp}
\begin{tabular}{@{}cccc@{}}
\toprule
\textbf{Temperature {[\unit{\celsius}]}} & Q-Q r-value & $T_D$ SD {[\unit{\ns}]} & Avg. Latency {[\unit{\us}]} \\ \midrule
\textbf{28}                 & 0.9292      & 197.1           & 1.931               \\
\textbf{40}                 & 0.9916      & 201.9           & 1.297               \\
\textbf{50}                 & 0.9928      & 242.2           & 1.051               \\
\textbf{60}                 & 0.0736      & 515.5           & 0.7749              \\ \bottomrule
\end{tabular}
\end{table}

\subsection{Model Uncertainty Estimation}
\label{sec:uncertainty}

\begin{table*}[ht]
\centering
\caption{Comparison to other work}
\label{tab:comparison}
\resizebox{\textwidth}{!}{%
\begin{tabular}{@{}rcccccc@{}}
\toprule
\multicolumn{1}{l}{}                      & \textbf{This Work}   & \textbf{\cite{dorrance2023energy}}           & \textbf{\cite{shukla2021ultralow}} & \textbf{\cite{cai2018vibnn}}      & \textbf{\cite{xu2021bayesian}}       & \textbf{\cite{fan2022fpga}}     \\ \midrule
\textbf{Implementation}                   & ASIC                 & ASIC                & Simulated & FPGA           & FPGA            & FPGA          \\
\textbf{Technology {[nm]}}                & 65                   & 22                  & 45 (PTM)  & 28 (Cyclone V) & 16 (ZU9EG)      & 20 (Arria 10) \\
\textbf{RNG} &
  \begin{tabular}[c]{@{}c@{}}Analog\\ (Thermal)\end{tabular} &
  TI-Hadamard &
  \begin{tabular}[c]{@{}c@{}}Analog\\ ($V_{th}$ Variation)\end{tabular} &
  Wallace &
  Box-Muller &
  MC Dropout \\
\textbf{Precision} &
  \begin{tabular}[c]{@{}c@{}}\texttt{INT8/4}\\ (Heterogeneous)\end{tabular} &
  \begin{tabular}[c]{@{}c@{}}\texttt{INT8/16/32}\\ \texttt{FP8/16/32}\\ \texttt{BF16}\end{tabular} &
  \texttt{INT4} &
  \texttt{INT8} &
  \texttt{INT16} &
  \texttt{INT8} \\
  \textbf{Area$^\mathrm{\ast}$ {[mm$^\mathrm{2}$]}}                      & \textbf{0.45}        & 3.88                & ---       & 80/17/100/39   & 2.9/1.4/6.6/8.6 & 71/52/97/86   \\
  \textbf{Norm. RNG Tput {[GSa/s/mm$^\mathrm{2}$]}} & \textbf{11.4 (62.3)$^\mathrm{\dag}$} & 1.20--1.88          & ---       & ---            & ---             & ---           \\
  \textbf{RNG Tput {[GSa/s]}}            & 5.12 \textbf{(28.0)$^\mathrm{\dag}$} & 4.65--7.31          & ---       & 13.63          & 8.88            & ---           \\
  \textbf{RNG Eff. {[pJ/Sa]}}            & \textbf{0.36}        & 1.08--1.69          & 0.37      & 38.8           & 5.40            & ---           \\
  \textbf{Norm. NN Tput {[GOp/s/mm$^\mathrm{2}$]}}  & 228 \textbf{(1246)$^\mathrm{\dag}$}  & 309--515            & ---       & ---            & ---             & ---           \\
  \textbf{NN Tput {[GOp/s]}}             & 102                  & \textbf{1200--2000} & ---       & 59.6           & ---             & 533--1590     \\
  \textbf{NN Eff. {[fJ/Op]}}             & 672                  & \textbf{31--65}              & ---       & ---            & --- & 24000--51000  \\ \bottomrule \vspace{1pt}
\end{tabular}%
}
{\raggedright $^\mathrm{\ast}$Percent utilization of LUT/Register/DSP/BRAM for FPGAs \\
$^\mathrm{\dag}$Scaled to \qty{22}{\nm} \par}
\end{table*}

\begin{figure}
    \centering
    \includegraphics[width=\columnwidth]{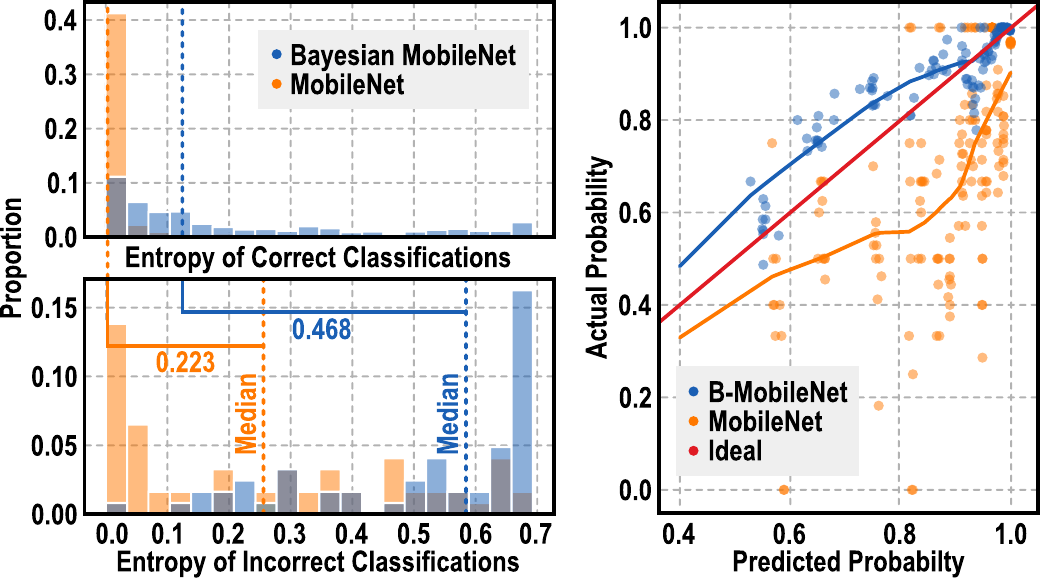}

	\caption{\textbf{Left)} BNNs significantly increase the entropy of incorrect
	and out-of-distribution classifications, providing more accurate uncertainty
	estimation. \textbf{Right)} Calibration curve showing low BNN ECE compared
	to the overconfident NN.}
    \label{fig:modelperf}

\end{figure}

\begin{figure}[h]
	\centering
	\includegraphics[width=\columnwidth]{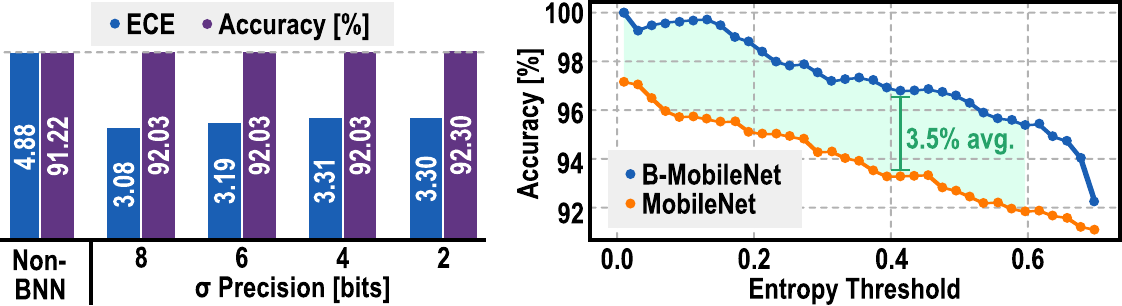}

	\caption{\textbf{Left)} Quantized partial-Bayesian MobileNet ECE and
	accuracy for the INRIA person dataset compared to a traditional network.
	\textbf{Right)} BNN vs NN accuracy recovery when removing high-entropy
	classifications. A lower threshold indicates more samples are removed.}
	\label{fig:accuracy}

\end{figure}

The INRIA person dataset \cite{dalal2005histograms} models a safety-critical
application where the neural network must accurately identify the presence of pedestrians.
MobileNet \cite{howard2017mobilenets} was chosen for this analysis due to its efficient feature
extraction capabilities, which lends itself to edge inference. 

Fig.~\ref{fig:modelperf} illustrates how BNNs achieve uncertainty estimation by
increasing the entropy of incorrect and out-of-distribution classifications. In
the standard MobileNet implementation, incorrect classifications typically
exhibit low entropy, making them indistinguishable \textit{a priori} from
correct classifications.  However, when employing a Bayesian classifier
(Bayesian FC layers, see Sec.~\ref{sec:arch}) on this chip, the the average
predictive entropy (APE) of incorrect classifications increases from 0.350 to
0.513 (+46.6\%).  Concurrently, the expected calibration error (ECE)
\cite{guo2017calibration} decreases from 4.88 to 3.31 (-32.2\%).  ECE measures
the total area between the ideal calibration curve (which is linear) and the
model's calibration curve; models with high ECE perform poorly at uncertainty
estimation, as demonstrated by the standard MobileNet in
Fig.~\ref{fig:modelperf}. 

Even with only 2 bits of $\sigma$ precision, the partial-Bayesian MobileNet
maintains a low ECE (see Fig.~\ref{fig:accuracy}).  However, this chip employs a
4-bit $\sigma$ to support more complex applications where the model requires
more precision to quantify uncertainty accurately.

In a real-world BNN model deployment, classifications with uncertainty exceeding
predetermined threshold would be flagged for further scrutiny, either via human
intervention or supplemental sensors and models. As demonstrated in
Fig.~\ref{fig:accuracy}, when deferring high-entropy classifications, the
partial-Bayesian network outperforms the standard model, achieving an average
accuracy recovery of $3.5\%$ for representative entropy thresholds between 0.0
and 0.6. Users can also trade-off overall accuracy for better uncertainty
estimation through adjustments to the loss function during training, although
such techniques are beyond the scope of this work.  

Overall, BNN models deployed on this chip effectively retain their uncertainty
estimation, confirming this chip's suitability for safety-critical edge
deployments, such as person detection for autonomous vehicles.

\section{Conclusion}
\label{sec:conclusion}

\begin{figure}
	\centering
	\includegraphics[width=0.8\columnwidth]{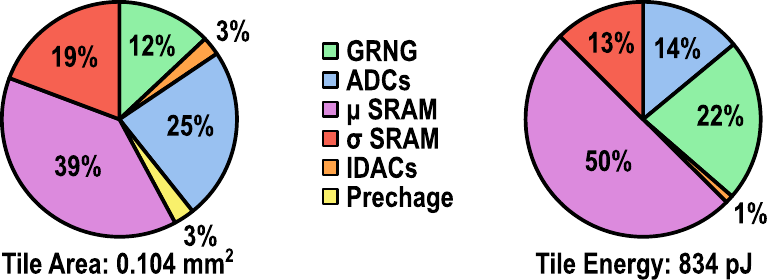}

	\caption{Tile area and energy breakdown for one complete MVM. Synthesized
	digital components, such as the calibration and reduction logic and IO
	buffers, are not included.}
	\label{fig:breakdown}

\end{figure}

Fig.~\ref{fig:breakdown} presents a detailed breakdown of the tile area and
energy consumption for one complete MVM. The SRAM accounts for over 63\% of the
total tile energy and 48\% of total tile area, underscoring the power and space
efficiency of the in-word GRNG cell. When compared to SOTA BNN ASICs, as
detailed in Tab.~\ref{tab:comparison}, this chip achieves a 75\% GRNG energy
reduction and increases GRNG throughput by over 6$\times$ per mm$^2$ at the
current tech node or over 33$\times$ per mm$^2$ when scaled to the same
technology.  

In conclusion, the integration of a \qty{360}{\femto\joule}/Sample GRNG directly
into SRAM memory words presents a significant advancement in the acceleration of
BNNs. By reducing the computational overhead associated with RNG and
facilitating fully-parallel CIM operations, this ASIC overcomes the traditional
challenges faced by BNN accelerators. The prototype chip validates the potential
of this approach to bring efficient AI uncertainty estimation to edge
computation without sacrificing model accuracy. This work paves the way for more
reliable and robust AI systems in safety-critical environments, ultimately
contributing to the broader adoption and implementation of BNNs in high-stakes
applications.

\bibliographystyle{IEEEtran}
\bibliography{references}

\end{document}